\documentclass{article}
\usepackage[utf8]{inputenc}
\usepackage[dvipsnames]{xcolor}
\usepackage{amssymb}
\usepackage{appendix}
\usepackage{graphicx}
\usepackage{mathtools}

\newcommand{\p}{\partial}
\newcommand{\e}{\mathrm{e}}
\newcommand{\sig}[1]{k({#1})}

\title{
	Measurement of a quantum system with a classical apparatus using ensembles on configuration space}
\author{Marcel Reginatto$^{1,}$\footnote{marcel.reginatto@ptb.de}\, and Sebastian Ulbricht$^1$}

\date{%
	$^1$\emph{Physikalisch–Technische Bundesanstalt, D–38116 Braunschweig, Germany}\\[1em] 
	(Dated: \today)
}

\begin{document}
	
	\maketitle
	
	\begin{abstract}
		Finding a physically consistent approach to modelling interactions between classical and quantum systems is a highly nontrivial task. While many proposals based on various mathematical formalisms have been made, most of these efforts run into difficulties of one sort or another. One of the first detailed descriptions was given by Sudarshan and his collaborators who, motivated by the measurement problem in quantum mechanics, proposed a Hilbert space formulation of classical-quantum interactions which made use of the Koopman-von Neumann description of classical systems. Here we use the approach of \textit{ensembles on configurations space} to give a detailed account of a classical apparatus measuring the position of a quantum particle that is prepared in a superposition of two localized states. We show that the probability of the pointer of the classical apparatus is left in a state that corresponds to the probability of the quantum particle. A subsequent observation of the pointer leads to an update of its probability density.
		From this we can obtain information about the position of the quantum particle, leading to an update of its wave function.
		Since this formalism incorporates uncertainties and finite measurement precision, it is well suited for metrological applications.
		Furthermore, it resolves fundamental issues that appear in the case of a quantum description of the apparatus.
	\end{abstract}
	
	
	\clearpage
	
	\section{Introduction}
	There are a number of good reasons for developing theoretical models that allow for interactions between quantum and classical systems. 
	%
	
	For many calculations, it is convenient or even necessary to describe parts of a complex system classically for physical or computational reasons, while the rest of the system is described quantum mechanically. 
	Ideally, such models should be free of inconsistencies and should allow for well defined approximation schemes. 
	%
	
	Mixed classical-quantum models also arise when questions of fundamental physics are considered. For some systems, e.g., such that involve  gravity, there is no full quantum theory yet. A detailed description of classical gravity (within the framework of general relativity) interacting with quantum matter (described by quantum field theory) can provide insight that helps in the search for a quantum theory of gravity. However, if gravity does not have to be quantized \cite{AlbersKieferReginatto2008,carlip_2008,Boughn_2009,Dyson2013,Rothman_2006}, it becomes essential to find a consistent theory that couples quantum fields to a classical gravitational field.
	%
	
	Another example from foundations of physics comes from measurement theory: in the Copenhagen interpretation of quantum mechanics, a measuring apparatus is described in classical terms \cite{Bohr1958,Omnes1999}. This implies a coupling (typically left undefined) between the quantum system and the classical apparatus. 
	%
	
	Due to the many applications in applied and fundamental physics,  various  approaches describing mixed classical-quantum systems have been developed. 
	The first general  model was worked out by E. C. G. Sudarshan \cite{Sudarshan1976}, who proposed a Hilbert space formulation incorporating the Koopman-von Neumann description of classical systems.	
	In this paper, we use instead the formalism of ensembles on configuration space \cite{HallReginatto2016,HallReginatto2005,Hall2008} to study the measurement of a quantum system with a classical apparatus.
	This approach brings to the fore a number of interesting conceptual issues, such as the question of how ``classicality'' may be defined, the role of the environment, and the connection between the state of the measuring apparatus and the collapse of the wave function \cite{HallReginatto2016}. 
	%
	
	The model discussed in this paper describes a simple type of measurement. However, the techniques that we use can also be applied to the description of complex measurement processes and provide a blueprint for modeling measurements carried out with more realistic devices.
	In particular, they allow us to consider the finite precision in the preparation of an initial quantum state and 
	the limitations on the calibration of the classical apparatus. Moreover, it is possible to keep track on the evolution of measurement uncertainties. This makes our approach highly suitable for the analytical discussion of non-idealized quantum particle position measurements and, thus, for modern high precision quantum metrology. 
	%
	
	Our paper is organized as follows. In order to present a consistent description of mixed classical-quantum systems, we first give an introduction to the formalism of ensembles on configuration space in Sec. \ref{sectionECS}. After that, in Sec. \ref{SectionModel}, we describe the measurement procedure in three steps:
	the  initial condition, the system during the interaction of the classical apparatus and the quantum particle, and the evolution of the classical pointer after the interaction. These steps are described in detail in the Secs. \ref{sec:interaction} and \ref{sectionPointerAfterInteraction}. Finally we consider the measurement of the classical pointer, from which the information about the  quantum particle position is obtained in Sec. \ref{sectionMeasurement}.
	A discussion of the implications of our analysis for the discussion of classical-quantum systems and for metrological application can be found in Sec. \ref{sec:discussion}.
	%
	
	Before we start with the main task of this publication, to describe the measurement of a quantum particle by a classical apparatus, we discuss briefly some of the difficulties of quantum measurement theory in the next section.
	%
	
	\section{Some aspects of the quantum measurement problem}
	\label{sectionQMP} 
	The earliest formal model of measurement goes back to von Neumann, who started with a fully quantum mechanical description of the measurement process and of both the quantum particle and the apparatus. He then went on to describe the unitary evolution of the quantum system that results when the coupling to the measuring apparatus is strong  \cite{vonNeumann1932}.
	The apparatus is left in a state that can be described by a reduced density matrix.
	In this formalism, however, difficulties arise when a macroscopic measurement apparatus is used: In this case, the device ends up with properties that are not compatible with observations \cite{BuschLahti2009,Holland1993,Omnes1999}. As Roland Omn\`es points out \cite{Omnes1999}, this issue arises due to the presence of off-diagonal elements of the density matrix: ``A linear superposition of macroscopically distinct states of the measuring apparatus is obtained'' and ``The theory predicts possible interferences between macroscopically different states of the apparatus that are associated with different data''. 
	%
	
	While there are a number of different proposals for dealing with this issue, there is no generally accepted resolution \cite{BuschLahti2009}. In von Neumann's original formulation, it is assumed that the unitary evolution of the quantum system in the moment of measurement is replaced  by a different type of evolution, the collapse of the wave function , that is discontinuous, noncausal and nonunitary.
	%
	
	In this paper, we consider a description of quantum measurement in which a \textit{classical} measuring apparatus is coupled to a quantum system using the formalism of ensembles on configurations space \cite{HallReginatto2016}. In this approach, the measurement of the quantum observable can be modeled via an interaction with a strictly classical measuring apparatus, followed by a direct measurement of the classical configuration. Our model of measurement is not subject to the problem discussed above, since the final state of the apparatus is described in terms of a classical mixture rather than a quantum superposition of macroscopic states. The measuring apparatus ends up in a particular state and is subject to a given uncertainty. This model supports the assumption of the Copenhagen interpretation that the pointer of a measurement device should be regarded as a truly `classical' object. %
	
	By extending the description of the measurement process to include the inevitable interaction of the classical measuring device with the environment, the ensemble that describes the apparatus and quantum system effectively decoheres into a proper mixture of noninteracting independent ensembles. 
	Thus, each element of the mixture after measurement satisfies the condition of \emph{strong separability},
	ruling out any possibility of nonlocal energy flow and nonlocal signaling \cite{HallReginatto2016}. It is important to point out that the role of the environment in this context differs from the one that we are accustomed to in decoherence theory \cite{JZKGKS2003}. Instead, it is analogous to the role that the environment plays in the description of  a cloud chamber track; that is, the environment selects a particular trajectory out of all the possible ones, which are consistent with the initial conditions \cite{Mott1929,Teta2009,FigariTeta2013}. 
	
	\section{Interacting classical-quantum systems: the approach of ensembles on configuration space}
	\label{sectionECS}
	Finding a physically consistent approach that models interactions between classical and quantum systems is a highly nontrivial task. 
	While many proposals based on various mathematical formalisms have been made, most of these efforts run into difficulties of one sort or another.
	One of the first detailed descriptions of mixed classical-quantum systems was given by Sudarshan and collaborators. 
	Motivated by the measurement problem in quantum mechanics, they proposed a Hilbert space formulation of classical-quantum interactions, which made use of the Koopman-von Neumann description of classical systems \cite{Sudarshan1976,Sudarshan1978,Sudarshan1979a,Sudarshan1979b}. In recent years, this work has been further developed and extended \cite{PhysRevA.63.022101,doi:10.1098/rspa.2018.0879,Gay_Balmaz_2020,PhysRevA.102.042221,Gay-Balmaz2021}.
	%
	
	Other common descriptions of classical-quantum interactions include the mean field approach, in which classical observables appear as parameters in a quantum Hamiltonian operator \cite{Boucher1988,Makri1999,Elze2012}, and the trajectory approach, which relies on modifying the de Broglie–Bohm formulation of quantum mechanics  \cite{Gindensperger2000,Prezhdo2001,Burghardt2004}. 
	In this paper, we will focus on an alternative description, the configuration ensemble approach, which is based on a general formalism for describing physical systems in terms of ensembles evolving on a configuration space \cite{HallReginatto2016}.
	While each of these approaches have proven useful for given applications (e.g., to provide an approximate description of a system where it is desirable to model some components classically while the rest of the system is modeled quantum mechanically), the question of their applicability in a more general sense still needs to be evaluated. 
	In particular, there is a lack of models that consider experiments in a realistic fashion.
	%
	
	Our description of interacting classical and quantum systems is formulated using ensembles on configuration space \cite{HallReginatto2016,HallReginatto2005,Hall2008}. 
	The basic idea of the approach is simple: physical systems are described by ensembles on configuration space, with  equations of motion derived from an action principle.
	The system is then described by an ensemble of configurations with probability density $P(x,t)$, where $P(x,t) \geq 0$ and $\int dx \,P(x,t) = 1 $. 
	To derive equations of motion, we introduce an \emph{ensemble Hamiltonian} $\tilde{H}[P,S]$, where $S$ is canonically conjugate to $P$. 
	In this Hamiltonian formulation, the evolution of the system is given by the equations
	\begin{equation}
		\frac{\partial P}{\partial t} = \left\{ P,\tilde{H} \right\}_{PB}
		= \frac{\delta\tilde{H}}{\delta S},~~~~~~~~~\frac{\partial S}{\partial t} = \left\{ S,\tilde{H} \right\}_{PB}
		=-\frac{\delta\tilde{H}}{\delta P}, \label{eqn:EOM}
	\end{equation}
	where $\{ A,B \}_{PB}$ is the Poisson bracket of two functionals of the canonical variables $P$ and $S$.
	%
	
	For the non-relativistic description of classical particle, the ensemble Hamiltonian is given by
	\begin{eqnarray}
		{H}_C[P,S] &=& \int dx\, P \left[ \frac{|\nabla S|^2}{2m} + V(x)\right] .\\ \nonumber
		{~}\label{HC} 
	\end{eqnarray}
	In order to describe the dynamics of a quantum particle,  an additional term is added to the classical ensemble Hamiltonian $H_C$:
	\begin{eqnarray}
		{H}_Q[P,S] &=& {H}_C[P,S]
		+  \frac{\hbar^2}{4} \int dx \,P\,\frac{1}{P^2}\frac{(\nabla P)^2}{2m} .\label{HQ} 
	\end{eqnarray}
	This additional term is proportional to the Fisher information of the probability density $P$ \cite{Fisher1925}. It gives rise to the term proportional to $\hbar^2$ that appears in Eq. (\ref{QEqMotion}), commonly known as the quantum potential due to the role assigned to it in Bohmian mechanics \cite{Holland1993}.
	%
	
	Evaluating the equations of motion of Eq.~(\ref{eqn:EOM}) using the quantum ensemble Hamiltonian ${H}_Q[P,S]$, we obtain
	\begin{equation}\label{QEqMotion}
		\frac{\partial P}{\partial t} + \nabla \cdot\left( P\frac{\nabla S}{m} \right) =0,~~~\frac{\partial S}{\partial t} + \frac{|\nabla S|^2}{2m} + V +  \frac{\hbar^2}{2m}\frac{\nabla^2 P^{1/2}}{P^{1/2}} = 0,
	\end{equation}
	which were first derived by Madelung in 1926 \cite{Madelung1926}. 
	In the case of $\hbar=0$, these equations  reduce to the classical equations of motion calculated from ${H}_C[P,S]$.
	The first equation in Eq. (\ref{QEqMotion}) is a continuity equation, the second equation is the classical Hamilton-Jacobi equation when $\hbar = 0$ and a modified Hamilton-Jacobi equation when $\hbar \neq 0$. 
	Introducing the complex canonical transformation $\psi:=\sqrt{P}~e^{iS/\hbar}$, $\psi^*:=\sqrt{P}~e^{-iS/\hbar}$, Eq. (\ref{QEqMotion}) takes the form of the Schr\"{o}dinger equation
	\begin{equation}
		i\hbar \frac{\partial \psi}{\partial t}
		= \frac{-\hbar^2}{2m}\nabla^2\psi + V\psi.
	\end{equation}
	%

	An advantage of the formalism introduced above is that it treats quantum and classical particles on an equal footing, only distinguished by the choice of ensemble Hamiltonian. 
	This feature allows the formalism to be extended in a natural and consistent way to mixed quantum-classical systems. 
	%
	
	A hybrid ensemble Hamiltonian where the classical and quantum particles interact via a potential term $V(q,x,t)$ is given by
	\begin{equation}\label{HQC}
		{H}_{QC}[P,S] = \int dq\,dx\, P\,\left[ \frac{|\nabla_x S|^2}{2M}
		+ \frac{|\nabla_q S|^2}{2m}+\frac{\hbar^2}{4P^2} \frac{|\nabla_q P|^2}{2m} + V(q,x,t)
		\right].
	\end{equation}
	Here $q$ denotes the configuration space coordinate of a quantum particle of mass $m$ and $x$ that of a classical particle of mass $M$. 
	The equations of motion for $P$ and $S$ derived from ${H}_{QC}$ are
	\begin{subequations}
		\begin{eqnarray}\label{EqsCQ}
			\frac{\partial P}{\partial t} &=&  -\nabla_q .\left( P
			\frac{\nabla_qS}{m} \right) - \nabla_x.\left(P\frac{\nabla_xS}{M}\right), \\ \nonumber
			{~} \\ 
			\frac{\partial S}{\partial t} &=& -
			\frac{|\nabla_qS|^2}{2m} - \frac{|\nabla_xS|^2}{2M} - V +
			\frac{\hbar^2}{2m}\frac{\nabla_q^2 P^{1/2}}{P^{1/2}} .
		\end{eqnarray}
	\end{subequations}
	In this way, mixed quantum-classical states are described in a unified way in terms of ensembles in configuration space, ensuring mathematical compatibility of the quantum and classical sectors.
	Furthermore, in this unified approach interactions between quantum and classical systems can be easily incorporated.
	%
	
	As one would expect, an ensemble for a system that consists of a classical and a quantum particle is defined to be \textit{independent} if the conditions
	\begin{equation} 
		P(x,q)=P_C(x)P_Q(q),\qquad \quad S(x,q)=S_C(x)+S_Q(q)
	\end{equation}
	are satisfied \cite{HallReginatto2016}. 
	We will assume this property in Sec. \ref{SectionModel} when defining initial conditions for a measurement process.
	%
	
	While the interpretation of $P$ as a probability density is straightforward, there are some subtle issues concerning its canonical conjugate $S$. 
	The physical interpretation of $S$ can be derived from its role in defining local energy and momentum densities. 
	To make this clear, we notice that the ensemble Hamiltonians satisfy the relation ${H}[\lambda P, S] = \lambda {H}[P,S]$. 
	Taking the derivative of this relation with respect to $\lambda$, using the Leibniz rule and evaluating the result at $\lambda=1$ leads to
	\begin{equation}\label{hodeg1} 
		{H} = \int dx \,P\frac{\delta {H}}{\delta P} = - \int
		dx\, P\frac{\partial S}{\partial t} = - \langle \partial S/\partial
		t \rangle ,
	\end{equation}
	which shows that $-\partial S/\partial t$ is a local energy density. 
	%
	
	Furthermore, one can check that $\int dx\, P\nabla S$ is the canonical infinitesimal generator of translations, since spatial variations of $P$ and $S$ can be expressed as
	\begin{subequations}
		\begin{eqnarray}
			\delta P(x) &=& \delta \textbf{x} \cdot \left \{ P, \int dx\,
			P\nabla S \right \}_{PB} = - \delta \textbf{x} \cdot \nabla P ,
			\\ 
			\delta S(x) &=& \delta \textbf{x} \cdot \left \{ S, \int dx\,
			P\nabla S \right \}_{PB} = - \delta \textbf{x} \cdot \nabla S ,
		\end{eqnarray}
		\label{eqb:translations_in_total}
	\end{subequations}
	under action of the generator. 
	Therefore $P\nabla S$ can be considered a local momentum density. These results are generally valid; i.e. they hold true for classical, quantum and mixed systems.
	%
	
	Having introduced the formalism to describe classical-quantum systems with ensembles in configuration space, we apply it in the next sections to the specific scenario of a classical apparatus measuring a quantum particle's position.
	%
	
	\section{Model of a quantum measurement with a classical apparatus}
	\label{SectionModel}
	Our aim is to describe a quantum measurement where the measurement apparatus is described classically using the approach of ensembles in configuration space, introduced in Sec. \ref{sectionECS}. 
	There we showed that the dynamics of the system and their interaction are given by ensemble Hamiltonians.
	Thus in this specific case we will be concerned with a Hamiltonian that has three contributions: the one describing the classical pointer, the one describing the quantum particle, and the one describing the interaction between the measuring apparatus and the quantum particle,
	\begin{subequations}
		\begin{eqnarray}
			H_C&=& \int d q d x\, P\left\{\frac{(\p_xS)^2}{2M}\right\},\\
			H_Q&=& \int d q d x\, P\left\{\frac{(\p_q S)^2}{2m}+\frac{\hbar^2}{8m}\frac{(\p_q P)^2}{P^2}\right\},\\
			H_{CQ}&=& \alpha(t) \int d q d x\, P\left\{\p_xS\, q\right\}, \label{eqn:interaction}
		\end{eqnarray}
	\end{subequations}
	where $\alpha(t)$ is the time dependent interaction strength. 
	This approach, already introduced in Ref.~\cite{HallReginatto2016}, follows the basic idea of von Neumann for the description of a strong measurement with the difference that the apparatus is classical.
	The interaction term, Eq.~(\ref{eqn:interaction}), couples the momentum of the classical pointer to the position of the quantum particle and only plays a role while the measuring device is turned on\footnote{In the Sudarshan model of mixed classical-quantum systems \cite{Sudarshan1976}, the interaction is given by the operator $\hat{H}_{CQ}=-i \alpha(t) q \frac{\partial}{\partial x}$ that is non-observable, since it contains the hidden variable $\hat{\lambda}_x=-i\frac{\partial}{\partial x}$. The functional $H_{CQ}$ of Eq. (\ref{eqn:interaction}), however, is an observable in the approach of ensembles on configuration space.}. 
	During this short time it is assumed to be strong enough that we can ignore the other contributions. 
	Thus we distinguish between three different phases of the measurement process: before, during, and after the interaction,
	\begin{subequations}
		\begin{eqnarray}
			H_1&=& H_C+H_Q \hspace{10em} t=0\\
			H_2&=& H_C+H_Q+H_{CQ}\,\,\approx\,\, H_{CQ} \hspace{2.5em}0<t\leq \varepsilon\\
			H_3&=& H_C+H_Q \hspace{10em} \varepsilon<t\leq t_m\,.
		\end{eqnarray}
	\end{subequations}
	The temporal sequence of this phases also resembles the sequence of the next sections:
	In Sec. \ref{sec:interaction} we will first introduce the initial condition at $t=0$ and then describe the system during the interaction phase ($t\leq \varepsilon$). The evolution of the classical pointer after the interaction  ($t>\varepsilon$) is the subject of Sec. \ref{sectionPointerAfterInteraction}. Finally we consider the measurement of the classical pointer at the time $t_m$ and the information that we obtain about the quantum system in Sec. \ref{sectionMeasurement}.
	%

	\section{Interaction between the classical measuring apparatus and the quantum system} \label{sec:interaction}
	We make the standard assumption that the state of the measuring apparatus does not depend on the state of the quantum system before the measurement takes place.
	This means that the correlations between the quantum system and the measuring apparatus only appear through the interaction during the measurement process.
	%
	
	We assume furthermore that the quantum system is isolated from the environment while the measurement takes place. 
	This necessarily brings in an asymmetry with respect to the measuring apparatus, which we assume satisfies some basic requirements of classicality \cite{HallReginatto2016}. 
	Classicality implies not only classical equations of motion but also that an  object that is described classically must be \emph{macroscopic}. 
	This is a reasonable assumption, since in the opposite case of \emph{microscopic} scales, objects are subject to quantum effects and any classical description must be given up in favor of a quantum mechanical treatment.
	%
	
	One important property of macroscopic objects is that they cannot be isolated from the environment: a macroscopic object cannot avoid scattering photons and other particles. 
	Thus, for example, the pointer of a measuring device will be continuously undergoing scattering processes, implying there are a large number of external degrees of freedom. 
	Another important property of macroscopic objects is that they have a very large number of internal degrees of freedom. 
	When describing a classical measuring apparatus, we commonly discuss only \emph{one} relevant degree of freedom, say the position of a pointer. 
	In reality, however, there is an enormous number of ``irrelevant'' degrees of freedom, say $10^{23}$, which must be treated statistically if they are modeled at all (usually they are simply neglected).
	Thus, classicality is not just an assumption about mathematical equations of motion, it is also an assumption about the physical model used for the description of a physical object.
	%

	\subsection{Initial conditions} \label{sec:initial}
	\begin{figure}[t]
		\centering
		\includegraphics[width=0.57\textwidth]{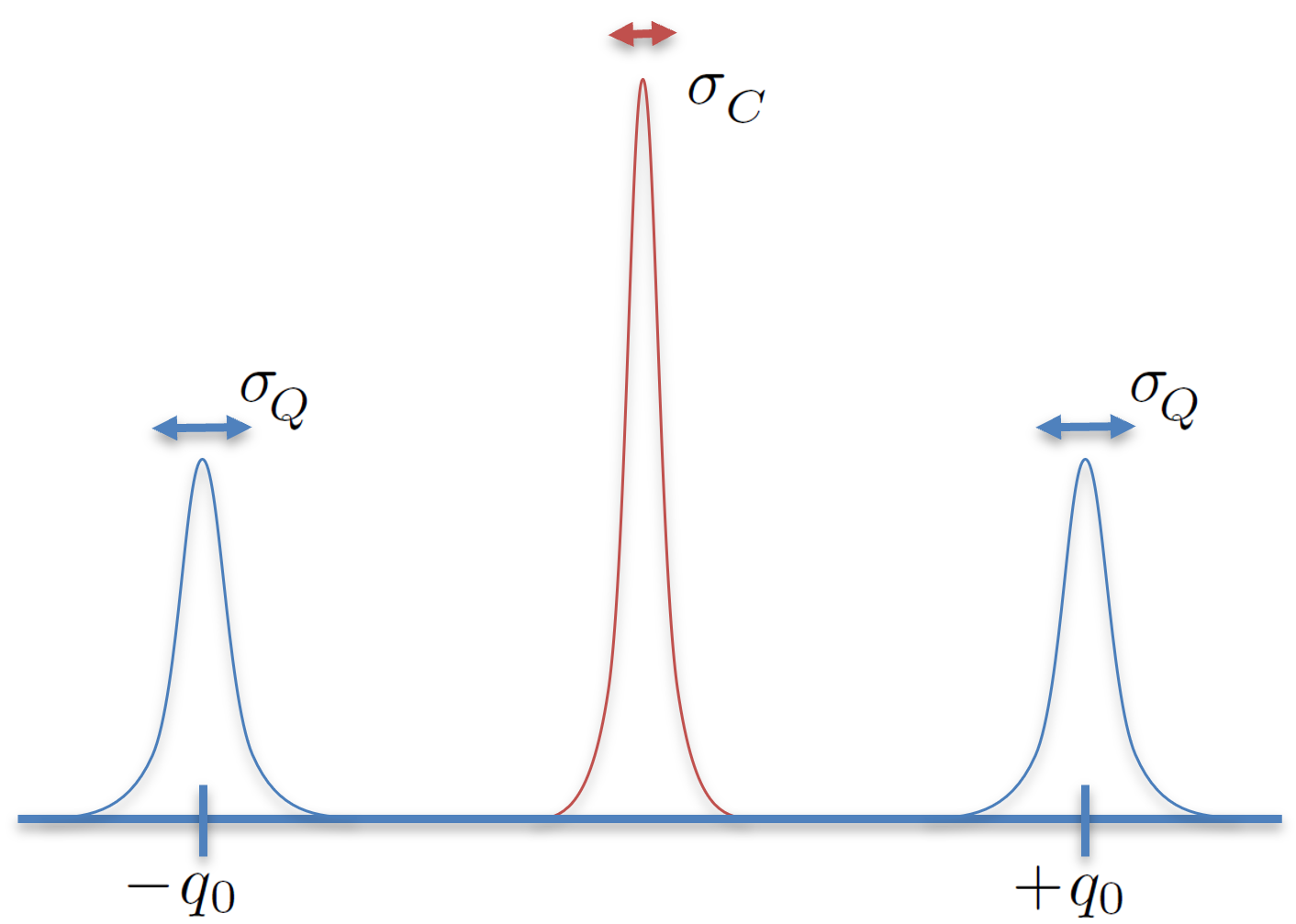}
		\caption{Initial condition for the probability density of the mixed quantum-classical system. The quantum particle is most likely localized around one of the two positions $q=\pm q_0$. The classical pointer is located in the center of the coordinate system. Both the classical pointer and the quantum system are initially prepared with finite precision, represented by the standard deviations $\sigma_C$ and $\sigma_Q$, respectively.} \label{fig:01}
	\end{figure}
	We assume that the two subsystems (classical and quantum) are independent at $t=0$. 
	The classical pointer's location is described by a Gaussian centered at $x=0$ and a standard deviation $\sigma_C$. 
	This allows for a description of measurements in which there is an initial uncertainty associated with the pointer of the classical measuring apparatus, which is unavoidable in a realistic description of a measurement. 
	The quantum system is in a superposition of two Gaussian wavefunctions localized at $q = \pm q_0$, each with a standard deviation $\sigma_Q$. 
	This describes a single quantum particle that is most likely localized around one of two positions.
	We consider the case where $\sigma_C<q_0$ so that the two peaks of the initial quantum probability density are well separated, see Figure \ref{fig:01}. 
	Thus, the initial probability density of the mixed classical-quantum system is given by
	\begin{eqnarray}
		P(x,q,0) &=& \frac{1}{\sqrt{2\pi}\,\sigma_C}\,\exp\left(-\frac{x^2}{2\sigma_C^2}\right)\nonumber \\
		&\qquad& \times \; \frac{1}{2\sqrt{2\pi} \,\sigma_Q} \left[\exp\left(-\frac{(q-q_0)^2}{2\sigma_Q^2}\right)+\exp\left(-\frac{(q+q_0)^2}{2\sigma_Q^2}\right) \right]\nonumber\\
		&=:&P_C(x,0;\sigma_C) \, P_Q(q,0;\sigma_Q,q_0) \label{eqn:initial_condition_P},
	\end{eqnarray}
	where we have chosen the prefactors, such that $P(x,q,0)$ is normalized to one.
	%
	
	A complete description of the initial conditions would also require a specification of the initial condition for the variable $S$. 
	However, it turns out that an explicit expression of $S(x,q,0)$ will not be needed for the rest of our analysis. 
	We will omit it here and refer to Appendix \ref{sec:ApproxInter} for those interested in more details concerning the initial state and the time evolution.
	%

	\subsection{The hybrid system during interaction ($0<t\leq\varepsilon$)} \label{sec:during_interaction}
	In our model we assume a strong measurement, where during the brief interaction time we can neglect the kinetic energy terms in the Hamiltonian. 
	During this time we only consider the changes in the state caused by the interaction term $H_{CQ}$. 
	Under this assumptions the equations of motion for $P$ and $S$ become
	\begin{equation} \label{eqn:EOM_approx}
		\frac{\p P}{\p t}=\frac{\delta H_{CQ}}{\delta S}\quad,\quad   \frac{\p S}{\p t}=-\frac{\delta H_{CQ}}{\delta P},
	\end{equation}
	which, given the explicit form of the interaction Hamiltonian Eq.~(\ref{eqn:interaction}), lead to the partial differential equations
	\begin{equation}
		\frac{\p P}{\p t}=-\alpha(t)\,q\,\p_x P\quad,\quad   \frac{\p S}{\p t}=-\alpha(t)\,q\,\p_x S\,.
	\end{equation}
	The general solution to this equations can be found by direct calculation and is given by
	\begin{equation}
		P(x,q,t)=P(x-q\sig{t},q,0)\quad  S(x,q,t)=S(x-q\sig{t},q,0)\,, \label{eqn:general_solution}
	\end{equation}
	where the dimensionless parameter $k(t)=\int_0^t dt' \, \alpha(t')$ is the integrated interaction strength.
	%
	
	As can be seen from Eq. (\ref{eqn:general_solution}), the interaction does not change the shapes of the functions $P$ and $S$ and only leads to the displacement $x \rightarrow x-q\sig{t}$. 
	This is consistent with the observation that $\int d x\, P\left\{\p_xS\right\}$ is the generator of translations of $x$, as shown by Eqs.~(\ref{eqb:translations_in_total}). 
	Also notice that, as a result of the interaction, the initially independent classical and quantum subsystems develop correlations and can not be treated independently anymore.
	%
	
	In what follows, we focus on the probability density that describes the location of the pointer at times $t=0$ and at $t=\varepsilon$; i.e., at the beginning and at the end of the interaction. 
	Initially, the probability density of the pointer is given by
	\begin{equation}
		P_C(x,0)=\int d q \, P(x,q,0)= \frac{1}{\sqrt{2\pi}\,\sigma_C}\,\exp\left(-\frac{x^2}{2\sigma_C^2}\right)\,.
	\end{equation}
	At the end of the interaction with the quantum system, however, the statistics of the pointer will be given by the integral of $P(x,q,t)$ from Eq.~(\ref{eqn:general_solution}) over the quantum coordinate $q$, evaluated at $t=\varepsilon$. 
	A direct calculation, which is presented in Appendix \ref{sec:ApproxInter}, leads to the probability density of the pointer in the form
	\begin{eqnarray}
		P_C(x,\varepsilon) &=& \int dq \, P(x,q,\varepsilon) \nonumber\\
		&=& \frac{1}{2 \sqrt{2\pi} \left(\sigma_C^2 + \sigma_Q^2 k^2(\varepsilon)\right)^{1/2}} \;  \left\{\exp\left[- \frac{\left(x - q_0 k(\varepsilon) \right)^2}{2 \left(\sigma_C^2 + \sigma_Q^2 k^2(\varepsilon)\right)}\right]\right. \nonumber\\ 
		&~& \qquad \qquad  \qquad \qquad + \left. \exp\left[- \frac{\left(x + q_0 k(\varepsilon) \right)^2}{2 \left(\sigma_C^2 + \sigma_Q^2 k^2(\varepsilon)\right)}\right]\right\}	\,. \label{eqn:P_C_epsilon}
	\end{eqnarray}
	As expected, this expression depends on the uncertainties of both the pointer and the quantum system. 
	%
	
	To interpret Eq.~(\ref{eqn:P_C_epsilon}) more easily, it is useful to consider the case where the initial uncertainty of the pointer $\sigma_C$ is small, such that the relation $\sigma_C << \sigma_Q \alpha(\varepsilon)$ holds. 
	This simplifies Eq.~(\ref{eqn:P_C_epsilon}) to
	\begin{eqnarray}
		P_C(x,\varepsilon;\sigma_C)\!\!\!
		&\rightarrow&\!\!\! \frac{1}{2\sqrt{2\pi} \sig{\varepsilon}\sigma_Q}\! \left\{\!\exp\!\left[-\frac{(x-\sig{\varepsilon}q_0)^2}{2(\sig{\varepsilon}\sigma_Q)^2}\right]\!\!+\!\exp\!\left[-\frac{(x+\sig{\varepsilon}q_0)^2}{2(\sig{\varepsilon}\sigma_Q)^2}\right]\! \right\}\,\,\,\nonumber\\[1em]
		\!\!\!&=&\!\!\! P_Q(x,0;\sig{\varepsilon}\sigma_Q,\sig{\varepsilon}q_0) \label{eqn:mimicQ}\,,
	\end{eqnarray}
	where the expression for $P_Q$ is given in Eq. (\ref{eqn:initial_condition_P}).  
	Thus we see that after interaction with the quantum system, \textit{the pointer ends up in a state with the same probability density as the quantum system at $t=0$, but with the location and standard deviation parameters scaled according to $\sigma_Q \rightarrow k(\varepsilon)\sigma_Q$ and $q_0 \rightarrow k(\varepsilon)q_0$}. 
	This is precisely the behaviour that allows us to use the classical pointer as  a measurement device for the quantum particle position.
	%
	
	\section{The evolution of the classical pointer after the interaction ends ($t>\varepsilon$)}\label{sectionPointerAfterInteraction}
	We are now interested in the evolution of the pointer after the interaction with the quantum system has come to an end. 
	To give an analytical description of the pointer at a time $t>\varepsilon$, we need to assume that the classical measuring device has been calibrated and that we know its response to the preceding interaction.
	This response of the apparatus is encoded in the function $\alpha(t)$ which we introduced in Eq. (\ref{eqn:interaction}) as the time-dependent strength of interaction, and now needs to be determined.
	%
	
	In what follows, we will consider an apparatus with a pointer that quickly responds to the interaction with the quantum system at the very beginning of the measuring process and moves with a constant velocity, afterwards.
	This assumption is reasonable, since it resembles the behaviour of detectors in real experiments.\footnote{To simplify the analysis, we will make one minor departure from the operating principles of measuring apparatuses: Measuring devices include damping forces that ensure that the pointer stops quickly once the apparatus is turned off. Without damping or friction, the pointer would continue at a constant velocity even after the interaction stops. We will assume here that there is \textit{no damping}, so that the pointer does indeed continue moving at a constant velocity after time $t=\varepsilon$. We are aware that this is an unrealistic assumption but it does simplify the analysis considerably. We ask the reader to consider then that this how our measuring apparatus works! (It would be rather straightforward to add a damping force to the analysis, but little would be gained by doing that.)}
	In our model, this means that the strength of the interaction $\alpha(t)=\lambda$ for $0< t<\varepsilon$ is a non-vanishing constant that depends on the characteristics of the detector and presumably needs to be  determined  by calibration. 
	This is equivalent to setting $\sig{t}=\lambda t$. As can be seen from Eq.~(\ref{eqn:general_solution}), this implies that the classical pointer moves with a constant velocity $\lambda q$ during the interaction time.
	%
	
	\subsection{Statistics of the classical pointer as the interaction comes to an end}
\begin{figure}[t] 
		\centering
		\includegraphics[width=\textwidth]{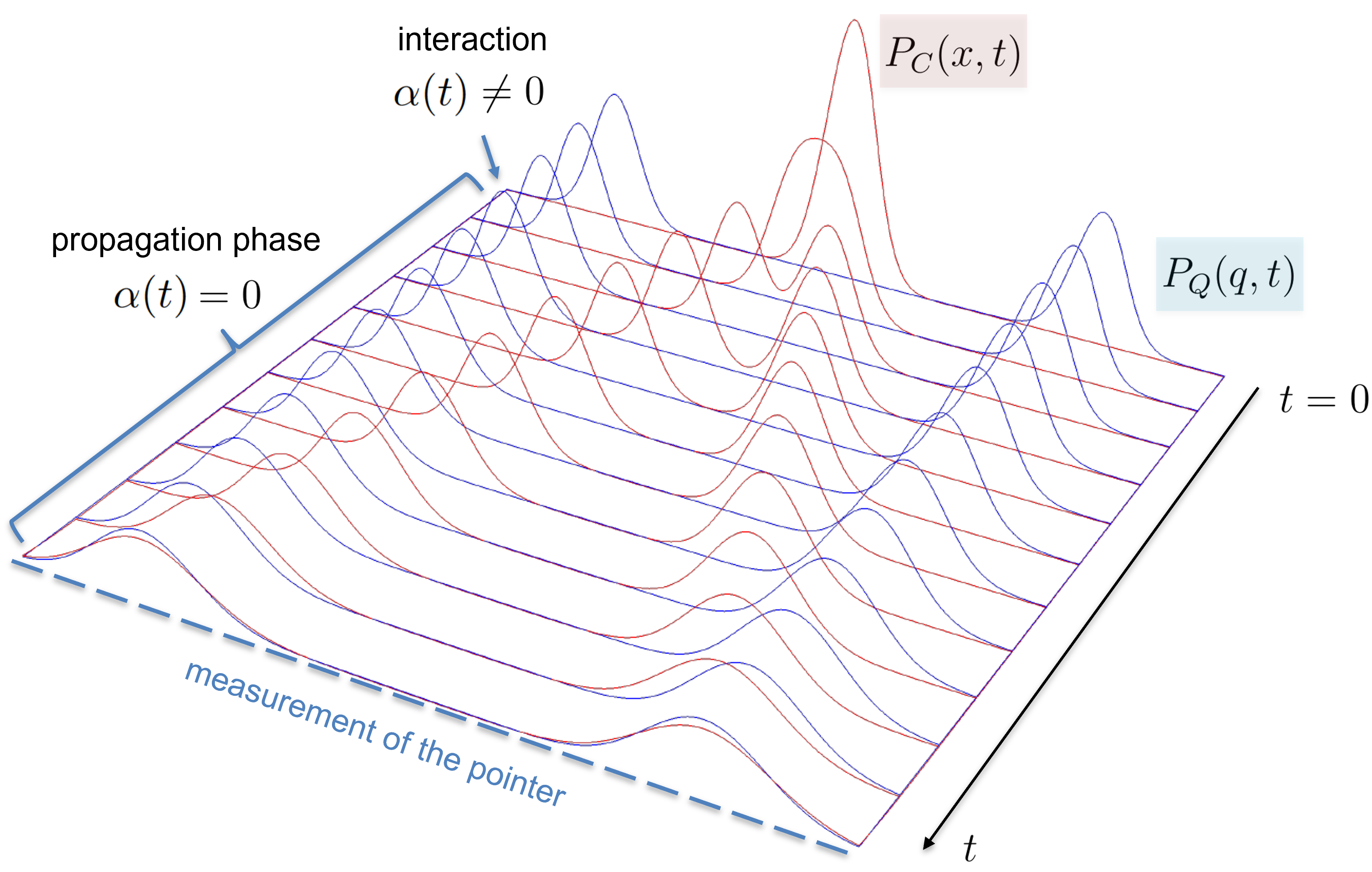}
		\caption{Time evolution of the classical (red) and the quantum subsystem (blue) after interaction. 
			Both systems propagate freely, after the interaction strength $\alpha(t)$, that determines the velocity of the pointer, is set to zero.
			One can see, that the initial Gaussian probability distribution for the pointer splits up and after some time resembles the bimodal distribution of the quantum system. This phase lasts until the pointer is measured and both subsystems are updated as described in Sec.~\ref{sectionMeasurement}. \label{fig:figure2}}
	\end{figure}

	Let us now consider the case where the initial uncertainty of the pointer $\sigma_C$ is small enough so that $\sigma_C << \sigma_Q \lambda t$. 
	Then, as the system approaches the time when the interaction is turned off, the statistics of the classical pointer is given by the probability density 
	\begin{eqnarray}
		P_C(x,t) &=& \frac{1}{2\sqrt{2\pi}\sigma_Q \lambda t} \exp\left[- \frac{\left(x - \lambda q_0  t \right)^2}{2\sigma_Q^2 \lambda^2 t^2}- \frac{\left(x + \lambda q_0  t \right)^2}{2\sigma_Q^2 \lambda^2 t^2}\right]. \qquad \label{Pxvarepsilon}
	\end{eqnarray} 
	This expression shows nicely how the initially Gaussian distribution for the classical pointer at $t=0$ splits up into two Gaussians that move with constant velocities $\pm \lambda q_0$ while they broaden with a standard deviation $\sigma_Q \lambda t$, as can be seen in Fig.~\ref{fig:figure2}.
	However, the probability density  Eq.~(\ref{Pxvarepsilon}) only contains partial information: 
	While it gives the correct statistics for measurements carried out on the pointer, it contains no information about correlations between $x$ and $q$. 
	For a complete description of these relations we would need to evolve the joint probability distribution $P(x,p,t)$ using the Hamiltonian equations of motion after choosing an appropriate function $S(x,q,t)$.
	Such an complete description, however, is not required, since after the interaction ceases, only the classical pointer is accessible.
	For times $t > \varepsilon$, both the classical pointer and the quantum particle independently obey the equations of motion of free particles.
	Thus, for our purposes it is enough to focus on the time development of the probability density $P_C(x,t)$ and how to extract information about the quantum particle position from a measurement of the pointer's motion.
	%

	\subsection{The motion of the pointer for $t > \varepsilon$: Two complementary descriptions} \label{sec:two_descriptions}
	In this section we will give two equivalent but complementary descriptions of the state of the pointer after the time of interaction, each associated with a particular interpretation of measurement. 
	The statistics of the pointer's position is defined in a unique way by the time dependent probability density $P_C(x,t)$. 
	We show, however, that it is possible to express this time dependent probability density in different ways using the concept of classical mixtures, which leads to different but equivalent descriptions of the state of the pointer corresponding to different types of measurements.
	%
	
	A brief review on classical mixtures is given in Appendix \ref{App:classicalMixtures}.
	In what follows, we will first show how the state of the pointer can be described as a classical mixture in terms of pairs of functions $P_{C|Q}(x,t;q)$, $S_{C|Q}(x,t;q)$ and a density for the parameter $q$, where $q$ is both a label for the elements of the classical mixture and the coordinate of the quantum particle. 
	Afterwards, we will discuss a second description, where the state is given in terms of a pair of functions $P_C(x,t)$, $S_C(x,t)$ without introducing a mixture of different elements.
	%

	\subsubsection{Description of the state of the pointer in terms of a classical mixture}
	As we have seen above, the assumption that the detector has a response $k(t)=\lambda t$ implies that the pointer moves with a constant velocity while the interaction takes place. 
	Since the position of the quantum particle is uncertain, the velocity of the pointer, given by $q\lambda$, will also be uncertain during the interaction and after the interaction ends. 
	The pointer then will  continue moving with a constant velocity $q\lambda$ that depends on the probability $P_{Q}(q,t)$ of finding the quantum particle particle at the location $q$.
	Thus, the statistics of the pointer after interaction can be formally described in terms of a \emph{classical mixture}. To introduce these mixtures, we use the product rule of probability theory $P(x,q,t)=P_{C|Q}(x|q,t)P_Q(q,t)$, which allows us to write a probability as the product of a conditional probability $P_{C|Q}(x|q,t)$ and a prior probability $P_Q(q,t)$. With this in hand, we can express the classical probability $P_C(x,t)$ by
	\begin{eqnarray}
		P_C(x,t) & = &\int d q \,P_{C|Q}(x|q,t)\,P_{Q}(q,t), \label{mix}
	\end{eqnarray}
	where each element of the mixture $P_{C|Q}(x|q,t)$ describes a classical particle moving with velocity $\lambda q$.
	To make this clear, we can bring $P_{C}(x,t)$ from Eq.~(\ref{Pxvarepsilon}) into the form
	\begin{eqnarray}
		P_C(x,t)   &=&\int d q \, \delta(x-\lambda qt)\nonumber\\
		&~& \times \frac{1}{2\sqrt{2\pi} \sigma_Q} \; \left[\exp\left(-\frac{(q + q_0)^2}{2 \sigma_Q^2}\right)+\exp\left(-\frac{(q - q_0)^2}{2 \sigma_Q^2}\right)\right], \label{eqn:mix2}
	\end{eqnarray}
	where we find $P_{C|Q}(x|q,t)=\delta(x-\lambda qt)$ in direct comparison with Eq.~(\ref{mix}).\footnote{The probability that the particle has velocity $\lambda q$} is given by the second line of Eq.~(\ref{eqn:mix2}). 
	By comparison with Eq.~(\ref{mix}) one could expect the appearance of a time dependent probability density for the propagating quantum particle. 
	The given expression, however, corresponds to the initial condition $P_Q(q,0;\sigma_Q,q_0)$ for the quantum particle from Sec. \ref{sec:initial}. 
	This is a consequence of the assumption that during the short time of interaction $0<t<\varepsilon$ the quantum system is ``frozen in time" and, therefore, is described by its initial condition.  
	In our model of a strong measurement this is valid, since the time $\varepsilon$ is considered to be smaller than any typical time scale of the dynamics of the quantum particle. 
	%
	
	To complete the description of the state of the pointer, we now consider the appropriate solutions of the Hamilton-Jacobi equation. 
	To each element of the mixture $P_{C|Q}(x|q,t)$ we can assign a corresponding function 
	\begin{equation}
		S_{C|Q}(x,t; q) = -\frac{M \lambda^2 q^2}{2}t + M \lambda q x \label{Scq}
	\end{equation}
	that is labeled by $q$. 
	This representation will be useful later when we consider a single measurement of the position of the pointer.
	%
	
	Notice that $S_{C|Q}(x,t; q)$ cannot be any arbitrary solution of the Hamilton-Jacobi equation, as the equations of motion of the ensemble have to be satisfied with the given choice of $P_{C|Q}(x|q,t)$. In Appendix \ref{App:classicalMixtures} we check that the functions $P_{C|Q}(x|q,t)$, $S_{C|Q}(x,t;\lambda q)$ satisfy the Hamilton-Jacobi equation and continuity equations
	\begin{equation}\label{QEqMotion2}
		\frac{\partial S_{C|Q}}{\partial t} + \frac{1}{2M}\left(\frac{\partial S_{C|Q}}{\partial x}\right)^2 = 0, \quad \frac{\partial P_{C|Q}}{\partial t} + \frac{\partial }{\partial x} \left( P_{C|Q}\frac{1}{M}\frac{\partial S_{C|Q}}{\partial x} \right) =0,\qquad 
	\end{equation}
	for an arbitrary choice of the parameter $q$.
	%
	
	In Sec.~\ref{sectionMeasurement} we will use the description of the state by classical mixtures to describe a measurement of the pointer that selects one of the elements of the mixture and leaves the pointer in that state. 
	Before we do that, however, we will briefly discuss the interpretation of the pointer's probability density, when no classical mixtures are introduced.
	%

	\subsubsection{Description of the state of the pointer without introducing a classical mixture}
	The statistics of the pointer as given by Eq. (\ref{Pxvarepsilon}) provides a second, equivalent description of the dynamics of the pointer after the interaction period. 
	This probability density $P_C(x,t)$ then describes two Gaussians travelling at constant velocity away from each other.  
	Each of the Gaussians has a width that changes linearly in time. 
	This is only possible if the velocity $v(x)=\frac{1}{M}\frac{\partial S}{\partial x}$ of the classical ensemble for a given time $t$ is proportional to $x$. 
	The corresponding function $S_C(x,t)$ that completes the description of the state of the pointer is given by
	\begin{equation} \label{SC}
		S_C(x,t) = \frac{Mx^2}{2t}.
	\end{equation}
	The functions $P_C(x,t)$ and $S_C(x,t)$ then satisfy both the Hamilton-Jacobi equation  and the corresponding continuity equation, 
	\begin{equation}\label{QEqMotion3}
		\frac{\partial P_C}{\partial t} + \frac{\partial }{\partial x} \left( P_C\frac{1}{M}\frac{\partial S_C}{\partial x} \right) =0,\qquad \frac{\partial S_C}{\partial t} + \frac{1}{2M}\left(\frac{\partial S_C}{\partial x}\right)^2 = 0,
	\end{equation}
	as can be checked by direct calculation. 
	This equivalent representation is useful if we would consider an experiment  to check the predicted probability density of the pointer by  a large number of measurements of identically prepared systems, without being concerned with a subsequent update of the pointer's state. 
	%
	
	\section{The measurement}
	\label{sectionMeasurement}
	The next and final step of our model is an observation of the pointer followed by an update of the probability densities of the pointer and the quantum particle. 
	We will also show that, unlike the usual description of a measurement, which assumes a quantum apparatus interacting with a quantum system, our approach does not rely on the linearity of quantum mechanics in order to compute the outcome of the measurement \cite{vonNeumann1932,BuschLahti2009}. 
	In this section, moreover, we will make use of the possibility of giving a description of the state of the pointer in terms of a classical mixture.
	
	\subsection{Measurements and update of the state of the pointer} \label{sec:update_the_state_of_pointer}
	If we carry out a large number of measurements of the pointer for an ensemble of identically prepared systems, we would expect the data to converge to the probability density of Eq. (\ref{Pxvarepsilon}). 
	However, this probability density does not describe the results we would expect when \emph{the same} pointer is measured repeatedly.
	The different statistics of the pointer in both cases were discussed in Sec.~\ref{sec:two_descriptions}.

	In what follows, we will consider a single measurement of the pointer \emph{at a particular time $t$}. To simplify the analysis, we furthermore neglect possible measurement uncertainties\footnote{The generalization to a non-ideal measurement is straightforward: A measurement with a given uncertainty will select a range of elements of the mixture, where the spread is determined by its uncertainty.}.
	Such a single, ideal measurement selects \textit{one} particular element $q'$ of the mixture in Eq. (\ref{mix}) from all the possible values $q$ that label the mixture. 
	This necessarily leads to an update of the state of the pointer according to
	\begin{equation}
		P_C(x,t) ~\rightarrow~ P_{C|Q}(x|q',t)=\delta(x-\lambda q't),\quad S_C(x,t) ~\rightarrow~ S_{C|Q}(x,t; q') \label{eqn:update}
	\end{equation}
	where $S_{C|Q}(x,t; q')$ is given by Eq. (\ref{Scq}) and $q'$ is a parameter that has now been determined by a measurement. 
	In this way, the measurement determines  the velocity of the pointer to be $v=\lambda q'$. 
	The model realistically predicts that the measured pointer (as a free particle) will stay in free motion, as it would be observed in such an experimental situation.
	%
	
	From the moment of the first measurement on, subsequent measurements will have to be consistent with the new probability density from Eq.~(\ref{eqn:update}). 
	Thus, we will never see the pointer randomly jumping to an other position, but moving smoothly in one direction along its trajectory $x(t)=\lambda q't$.
	%
	
	As a result of the update of Eq.~(\ref{eqn:update}), the state of the pointer no longer depends on the configuration space coordinate $q$ of the quantum particle. 
	Instead, the measured quantity $q'$ enters the equations. 
	Thus, we find that the hybrid classical-quantum ensemble decomposes into \emph{independent} classical and quantum subsystems.  
	This rules out any possibility of nonlocal energy flow and non local signaling between classical and quantum subsystems \cite{HallReginattoSavage2012}, as discussed in more detail in Ref. \cite{HallReginatto2016}.
	%
	
	\subsection{The role of the environment}
	Before we describe in the next section how a measurement of the classical pointer leads to an update of the quantum system, it lends itself to briefly discuss how the classical pointer is actually measured. 
	Generally, there are many conceivable ways of obtaining information about the pointer's position. 
	One way to obtain this information is to shine light on the pointer.
	Then the environment is filled with particles, particularly  photons in this case. 
	These particles will be scattered by the pointer at its current position.
	This way, an observer ``looking at the pointer" or camera taking a picture, would obtain the information about the pointer's position from the scattered light.
	Under normal conditions, a large number of photons will scatter from the pointer while a measurement takes place, and each of these scattering events may be considered a potential measurement. 
	%
	
	While the environment plays an important rule in the measurement process, it does not affect the equations that describe the motion of the pointer: 
	Since the photons are incident from all directions, the average transfer of momentum will be null. Moreover, the difference in momenta between the macroscopic pointer and a single photon is so large that one can neglect the recoil of the pointer as it scatters the photons.
	%
	
	The consequences of the coupling between the classical pointer and the environment can be described by an approach \cite{HallReginatto2016} similar to that by N. F. Mott in 1929, explaining the appearance of \textit{straight tracks} in a Wilson cloud chamber when an atom undergoes radioactive decay with the emission of an $\alpha$-particle \cite{Mott1929,Heisenberg1930}:
	While the $\alpha$-particle is initially described as a spherical matter wave that emerges from the atomic nucleus, the continuous interaction with the environment picks out a particular direction of motion.
	%
	
	The technical details of our analysis of the measurement of the classical pointer are of course very different from the ones in Mott's argument, but the analogy is clear: in our case, an initial probability density that describes a mixture of pointer states with different speeds and directions of motion is updated due to the interaction with the environment. Consequentially, all scattering events in which photons scatter from the pointer can only take place along the trajectory of one particular element of the mixture.
	%
	
	\subsection{Update of the state of the quantum system}  \label{sec:FreeQuantumParticle}
	Having discussed the update of the state of the classical pointer in the previous sections,  we can now discuss the update of the state of the quantum system, which is the last step in the measurement process. 
	We  consider again the case where the initial uncertainty of the pointer $\sigma_C$ is small enough so that $\sigma_C << \sigma_Q \lambda t$,  as already assumed in Sec.~\ref{sec:during_interaction}.  
	Under this assumption, the joint probability of the mixed clasical-quantum system can be expressed in the form
	\begin{eqnarray}
		P(x,q,t) &=:&  P_{Q|C}(q|x,t)P_C(x,t) \nonumber\\
		&=&  \frac{\lambda t}{\sqrt{2\pi}\sigma_C}\exp\left[-\frac{ \lambda^2 t^2 }{2 \sigma_C^2} \left(q - \frac{ x}{ \lambda t }\right)^2 \right]\nonumber\\
		&~& \times \frac{1}{2\sqrt{2\pi}\sigma_Q \lambda t} \exp\left[- \frac{\left(x - \lambda q_0  t \right)^2}{2\sigma_Q^2 \lambda^2 t^2}- \frac{\left(x + \lambda q_0  t \right)^2}{2\sigma_Q^2 \lambda^2 t^2}\right]\,, \label{eqn:joint_probability}	
	\end{eqnarray}
	where in the first equality we have used the product rule of probability theory to express $P(x,q,t)$ in terms of the conditional probability $P_{Q|C}(q|x,t)$ and the prior probability $P_C(x,t)$. 
	A detailed calculation of Eq.~(\ref{eqn:joint_probability}) can be found in Appendix \ref{sec:ApproxInter}
	%
	
	We suppose now that at time $t=t_{m}$ we measure the position $x=x_{m}$ of the pointer. 
	For the sake of simplicity we will assume a measurement of $x_m$ and $t_m$ with negligible uncertainty\footnote{As already mentioned in Sec.~\ref{sec:update_the_state_of_pointer} the generalization to a non-idealized measurement would be straightforward.}. 
	Then the probability density for the location of the pointer has to be updated to a delta function, and the probability for $q$ has to be updated to 
	\begin{equation}
		P_{Q|C}(q|x_{m},t_{m}) =  \frac{1}{\sqrt{2\pi}\left(\frac{\sigma_C}{\lambda t_{m}}\right)}\exp\left[-\frac{ 1 }{2 \left(\frac{\sigma_C}{\lambda t_{m}}\right)^2} \left(q - \frac{ x_m}{ \lambda t_m }\right)^2 \right] \,.
	\end{equation}
	This means that the measurement localizes $q \rightarrow q_m$ with an uncertainty that depends on the initial uncertainty of the pointer  according to
	\begin{equation}
		q_m \sim \frac{ x_m}{ \lambda t_m }, \qquad \sigma_{Q,m} = \frac{\sigma_C}{\lambda t_{m}}.
	\end{equation}
	This expressions are independent of the  uncertainty of the initially prepared quantum system, as we would expect from such a measurement. 
	Since the equations of motion for the quantum particle have to be fulfilled also after measurement, the update will affect both variables $P$ and $S$. This leads to what is described as the \emph{collapse of the wave function} in the literature on quantum measurement theory. 
	%
	
	Notice that the probability density for the quantum particle does not depend on the configuration space coordinate $x$ of the classical pointer, but only on the measured value $x_m$. As pointed out above, in the discussion on the update of the state of the pointer, the hybrid classical-quantum ensemble decomposes into independent classical and quantum subsystems.
	%
	
	\section{Summary and Discussion} \label{sec:discussion}
	In this work we have used the approach of ensembles on configuration space to construct a model where a \textit{classical} apparatus (in the precise sense given throughout the previous sections) is used to measure a quantum system. In Sec. \ref{sectionECS} we start with the introduction of the the mathematical formalism of this approach to mixed classical-quantum systems. The measurement process was then described in Sec. \ref{SectionModel} - \ref{sectionPointerAfterInteraction} in three chronological steps:
	the  initial condition, the system during the interaction of the classical apparatus and the quantum particle, and the evolution of the classical pointer after the interaction. Finally, in Sec. \ref{sectionMeasurement}, we discussed the measurement of the pointer of the classical apparatus, from which the information about the  position of the quantum particle was obtained.
	
	We have shown that describing the apparatus classically within the formalism of  ensembles on configuration space solves the main difficulties, discussed in Sec.  \ref{sectionQMP}, which would arise when also the apparatus is described quantum mechanically. 
	One important feature of the model is the asymmetry between the measuring apparatus and the quantum system being measured: While we assume that the quantum system can be isolated from the environment during the measurement,  the classical system inevitably interacts with the environment. As a consequence, the environment forces an update of the probability density of the pointer. This way, the classical mixture which mirrors the probability density of the quantum state is updated to a single element of the  mixture. 
	The motion of the pointer is then described by this very element, without the possibility of being changed to another element by additional measurements.
	A further, more technical, difference between the treatment of classical and quantum systems is that one may describe the quantum system in terms of a superposition of eigenstates (as per the linearity of quantum mechanics), while this is not possible for the state of the pointer: 
	Since after the measurement one element is chosen out of the classical \textit{mixture}, it is
	impossible for the pointer to be left in a state where there is interferences between macroscopically different states. This is of advantage in comparison to the purely quantum description of the measuring process, as discussed in Sec. \ref{sectionQMP}. Furthermore, the interpretation of the probabilities that appear in mixtures are straightforward as they are clearly epistemic in nature.
	
	One can conclude that the notion of the ``classicality'' of the measuring apparatus requires more than just classical equations of motion for the pointer. In addition, it is essential that the system has to be macroscopic, can not be decoupled from the environment, and has a large number of degrees of freedom in order to describe a system as ``classical''.
	
	In our model, the collapse of the wavefunction of the quantum system corresponds to an update which takes place when the classical pointer is measured. As a result of this update, the mixed classical-quantum ensemble decomposes into two independent subsystems, classical and quantum, which are no longer correlated. This also prevents any superluminal communication between the pointer and the quantum system \cite{HallReginattoSavage2012}, as discussed in more detail in Ref. \cite{HallReginatto2016}.

	Since our model allows for a detailed description of the steps of the measurement process, it is possible to keeps tracks of the uncertainties associated with each step. In particular, they allow us to consider the finite precision in the preparation of an initial quantum state and the limitations on the calibration of the classical apparatus.
	Therefore, our approach enables the analytical discussion of non-idealized quantum particle position measurements and is applicable to the challenges of modern high precision quantum metrology. The comparatively simple model we used to introduce the basic formalism can be extended to describe even more complex measurement processes.

	\section*{Acknowledgement}
	We want to thank Michael J.
	W. Hall for comments and valuable discussions. We also want to thank the organizers of the conference Koopman Methods in Classical and Classical-Quantum Mechanics for a very stimulating meeting. One of the authors acknowledges support by the Deutsche
	Forschungsgemeinschaft (DFG, German Research
	Foundation) under Germanys Excellence Strategy - 
	EXC-2123 QuantumFrontiers - 390837967.
	
	\appendix
	\addcontentsline{toc}{section}{Appendices}
	\section*{Appendices}

	\section{Validity of the approximation during the time of interaction}
	\label{sec:ApproxInter}
	
	In Sec.~\ref{sec:during_interaction} we have introduced an approximate solution of the modified Hamilton-Jacobi equation and continuity equation during the time of interaction under the assumption of a strong coupling constant. We made the further assumption that the interaction is effectively instantaneous. In what follows, we demonstrate the validity of this approximation under the additional assumption that the pointer is originally at rest.
	
	\subsection{Initial state for the quantum particle}
	
	The wavefunction that we use as initial condition is the sum of two stationary Gaussian wavefunctions which are located at $\pm q_0$. The full wave function of the quantum system can be written as
	\begin{eqnarray}
		\psi(q,t) &=& \frac{1}{\sqrt{2}}[\psi_+(q,t) + i\psi_-(q,t)],\\
		\psi_{\pm}(q,t)&=&\sqrt{P_{\pm}(q,t)}\, \e^{iS_{\pm}(q,t)/\hbar} \label{eqn:quantum_system}
	\end{eqnarray}
	where \cite{Holland1993}
	\begin{eqnarray}
		P_Q^{\pm}(q,t) &=& \frac{1}{(2\pi \sigma^2(t))^{1/2}}  \exp\left[-\frac{1}{2}\left(\frac{q \pm q_0}{ \sigma(t)}\right)^2\right],\\
		S_Q^{\pm}(q,t) &=& -\frac{\hbar}{2}\tan^{-1}\left(\frac{\hbar t}{2m \sigma_Q^2}\right) + \frac{(q \pm q_0)^2 \hbar^2 t}{8 m \sigma_Q^2 \sigma^2(t)},\\
		\sigma(t) &=& \sigma_Q\left[1+\left(\frac{\hbar t}{2 m \sigma_Q^2}\right)^2\right]^{1/2}.
	\end{eqnarray}
	In particular, we need $P_Q$ and $S_Q$ at time $t=0$ to be able to calculate the evolution of the classical pointer after interaction. Since $S_{\pm}(q,0)=0$, the wavefunction reduces to $\psi(q,0) = \sqrt{P_Q^+(q,0)} + i\sqrt{P_Q^-(q,0)}$, so that we have
	\begin{eqnarray}
		P_Q(q,0) &=& \frac{1}{2\sqrt{2\pi }\,\sigma_Q}  \left[\exp\left(-\frac{(q + q_0)^2}{2 \sigma_Q^2}\right)+\exp\left(-\frac{(q - q_0)^2}{2 \sigma_Q^2}\right)\right],\\
		S_Q(q,0) &=& 0.
	\end{eqnarray}
	
	\subsection{Initial state for the classical pointer}
	
	We assume that the initial state of the pointer is described by a \emph{classical mixture} (see Appendix \ref{App:classicalMixtures} for a definition). The elements of this mixture consist of states labeled by a parameter $v$, of the form
	\begin{eqnarray}
		S_{C|v}(x|v,t) &=& -\frac{Mv^2t}{2} + Mvx,\\
		P_{C|v}(x|v,t) &=& \frac{1}{\sqrt{2\pi}\, \sigma_C} \; \exp\left(-\frac{(x - vt)^2}{2 \sigma_C^2}\right)\,.
	\end{eqnarray}
	We have introduced the notation ``$|v$'' to emphasize that these quantities are defined with respect to the parameter $v$, which corresponds to the velocity of the pointer, as can be seen by evaluating $\frac{1}{M}\frac{\partial  S}{\partial x}=v$. The probability density $\rho(x,p,t)$ associated with this classical mixture is given by
	\begin{equation}
		\rho(x,p,t) =\int \, d v \,  P_v(v)  \, \delta\left(p-\frac{\partial  S_{C|v}(x|v,t)}{\partial x}\right) \,  P_{C|v}(x|v,t) 
	\end{equation}
	where $ P_v(v)  \ge 0$ and $\int \, dv \,  P_v(v) =1$ so that $\int dx\, dp\,\rho(x,p,v)=1$. One can show that such a classical mixture is strictly equivalent to a classical phase-space distribution \cite{HallReginatto2016}.
	
	For the purpose of checking the approximation, we do not need to specify the details of the mixture such as the functional form of $P_v(v)$. We will only assume that such a distribution is symmetric and the mixture satisfies $<v>=0$ so that the pointer is stationary. At the end of the calculation, we will consider the limit where the velocity of the pointer goes to zero.
	
	\subsection{The approximate solution}
	It is not necessary to use the full mixture when examining the validity of the approximate solution. It is sufficient to consider any particular element 
	$S_{C|v}(x|v,t)$, $P_{C|v}(x|v,t)$.
	We consider the initial conditions 
	\begin{eqnarray}
		S(x,q,0) &=&  S_{C|v}(x|v,t)  + S_Q(q,0) 
		= Mvx,\\[1em]
		P(x,q,0) &=&  P_{C|v}(x|v,t)  P_Q(q,0)  \nonumber\\
		&=& \frac{1}{4\pi \sigma_C \sigma_Q} \; \exp\left(-\frac{(x-vt)^2}{2 \sigma_C^2}\right)\nonumber\\
		&~& \qquad\times \left[\exp\left(-\frac{(q + q_0)^2}{2 \sigma_Q^2}\right)+\exp\left(-\frac{(q - q_0)^2}{2 \sigma_Q^2}\right)\right]
	\end{eqnarray}
	for the combined system of quantum particle and classical pointer.
	
	As in Sec.~\ref{sectionPointerAfterInteraction}, we assume an interaction strength of the form $\alpha(t)=\lambda q t$ during the time $0 < t < \varepsilon$. Under this assumptions, we obtain
	\begin{eqnarray}
		S(x,q,0 < t < \varepsilon)  &=&  S(x-\lambda q t,q,0) \\
		P(x,q,0 < t < \varepsilon)  &=&  P(x-\lambda q t,q,0) 
	\end{eqnarray}
	Then, during $0 < t < \varepsilon$, the explicit form of $P$ and $S$ is given by
	\begin{eqnarray}
		S(x,q,t) &=& Mv(x-\lambda q t)\label{AppSv}\\[1em]
		P(x,q,t) &=& \frac{1}{4\pi \sigma_C \sigma_Q} \; \exp\left(-\frac{(x-vt-\lambda q t)^2}{2 \sigma_C^2}\right)\nonumber\\
		&~& \qquad\times \left[\exp\left(-\frac{(q + q_0)^2}{2 \sigma_Q^2}\right)+\exp\left(-\frac{(q - q_0)^2}{2 \sigma_Q^2}\right)\right]\label{AppPv}
	\end{eqnarray}
	
	If we now introduce again the kinetic energy terms of the Hamiltonian that we previously neglected because the interaction term was dominant, the modified Hamilton-Jacobi equation is given by
	\begin{eqnarray}
		-\frac{\partial S}{\partial t} &=& \left[\frac{1}{2M}\left(\frac{\partial S}{\partial x}\right)^2 + \frac{1}{2m}\left(\frac{\partial S}{\partial q}\right)^2 + \frac{\hbar^2}{2m}\frac{1}{\sqrt{P}}\frac{\partial^2 \sqrt{P}}{\partial q^2}\right] + \lambda q\frac{\partial S}{\partial x}\nonumber\\
		\Rightarrow -Mv\lambda q 
		&=& \left[ \frac{Mv^2}{2} + \frac{M^2v^2\lambda^2 t}{2m} + \frac{\hbar^2}{2m}\left(\frac{1}{\sqrt{P}}\frac{\partial^2 \sqrt{P}}{\partial q^2}\right)  \right]  -Mv\lambda q
	\end{eqnarray}
	where the terms in square brackets are the ones that we previously omitted.
	
	The third term in square brackets is the quantum potential term. One can show that it remains finite provided both $\sigma_C, \sigma_Q > 0$. Furthermore, since the quantum potential is the only term that is proportional to $\hbar^2$, we can neglect it with respect to the other ``classical'' terms (without $\hbar$) that contribute to the energy. Finally, in the limit $v \rightarrow 0$, we can neglect all remaining terms of order $v^2$ when compared to the terms linear in $v$. Thus all the terms in the square brackets can be neglected. This shows that the approximation is indeed valid. 
	
	One must also check that the continuity equation,
	\begin{equation}
		\frac{\partial P}{\partial t} + \frac{\partial }{\partial x}\left( P\frac{1}{M}\frac{\partial S}{\partial x} \right) + \frac{\partial }{\partial q}\left( P\frac{1}{m}\frac{\partial S}{\partial q} \right) = 0,
	\end{equation}
	is also valid during the time of interaction ($0 < t < \varepsilon$) under the assumptions of the approximation. A straightforward calculation using Eqs. (\ref{AppSv}-\ref{AppPv}) shows that the equation is indeed satisfied provided all terms of higher order in $v$ and $t$ are neglected with respect to the terms linear in $v$; that is, in the limit $v\rightarrow 0$ and for the case of small interaction times $\varepsilon$. 
	
	\section{Explicit expressions for probability densities}
	\label{sec:ApproxInter}
	In what follows, for the sake of completeness, we give some explicit forms for probability densities that have been used for the calculation.
	We begin with the initial probability density, given by
	\begin{eqnarray}
		P(x,q,0) &=& P_C(x,0) P_Q(q,0)  \nonumber\\
		&=& \frac{1}{4\pi \sigma_C \sigma_Q} \; \exp\left(-\frac{x^2}{2 \sigma_C^2}\right)\left[\exp\left(-\frac{(q + q_0)^2}{2 \sigma_Q^2}\right)\right. \nonumber\\
		&~& \qquad \qquad \qquad \qquad \qquad \qquad \left.+\exp\left(-\frac{(q - q_0)^2}{2 \sigma_Q^2}\right)\right]\nonumber\\
		&=& \frac{\sqrt{\eta_C \eta_Q}}{4\pi} \; \exp\left(-\frac{\eta_C x^2}{2 }\right) \left[\exp\left(-\frac{\eta_Q(q + q_0)^2}{2}\right) \right. \nonumber\\
		&~& \qquad \qquad \qquad \qquad \qquad \qquad \left.+ \exp\left(-\frac{\eta_Q(q - q_0)^2}{2} \right)\right] \nonumber\\
		&=& \frac{\sqrt{\eta_C \eta_Q}}{4\pi} \;  \left[\exp\left(-\frac{\eta_Q(q + q_0)^2+\eta_C x^2}{2}\right) \right. \nonumber\\
		&~& \qquad \qquad \qquad \qquad \left. + \exp\left(-\frac{\eta_Q(q - q_0)^2+\eta_C x^2}{2} \right)\right]
	\end{eqnarray}
	where we introduced the \textit{precision} $\eta := 1/\sigma^2$ to simplify the formulae. 
	
	At any time $t$ during the interaction, we have 
	\begin{equation}
		P(x,q,\epsilon) = P(x-q k(t),q,0), 
	\end{equation}
	and the probability density can be written in a form that is  convenient for calculating the statistics of the pointer by integrating with respect to $q$, 
	\small
	\begin{eqnarray}
		P(x,q,t) &=& \frac{\sqrt{\eta_C \eta_Q}}{4\pi} \;  \left[\exp\left(-\frac{\eta_Q(q + q_0)^2+\eta_C (x-q k(t))^2}{2}\right)\right. \nonumber\\ 
		&~& \qquad \qquad +  \left. \exp\left(-\frac{\eta_Q(q - q_0)^2+\eta_C (x-q k(t))^2}{2} \right)\right]\nonumber\\
		&=& \frac{\sqrt{\eta_C \eta_Q}}{4\pi} \;  \left\{\exp\left[-\frac{\eta_Q + \eta_C k^2(t) }{2} \left(q + \frac{\eta_Q q_0-\eta_C k(t) x}{\eta_Q + \eta_C k^2(t) }\right)^2 \right. \right. \nonumber\\
		&~& \qquad \qquad \qquad \qquad \qquad \qquad \qquad \left. \left. - \frac{\left(x - q_0 k(t) \right)^2}{2\sigma_C^2 \left(1 + \sigma_Q^2 k^2(t)/\sigma_C^2\right)}\right]\right. \nonumber\\ 
		&~& \qquad \qquad +  \left. \exp\left[-\frac{\eta_Q + \eta_C k^2(t) }{2} \left(q - \frac{\eta_Q q_0+\eta_C k(t) x}{\eta_Q + \eta_C k^2(t) }\right)^2 \right. \right. \nonumber\\
		&~& \qquad \qquad \qquad \qquad \qquad \qquad \qquad \left. \left. - \frac{\left(x + q_0 k(t) \right)^2}{2\sigma_C^2 \left(1 + \sigma_Q^2 k^2(t)\right/\sigma_C^2)}\right]\right\}\nonumber\\
		&=& \frac{1}{4\pi \sigma_C \sigma_Q} \;  \left\{\exp\left[-\frac{ \left(1 + \sigma_Q^2 k^2(t)/\sigma_C^2 \right) }{2 \sigma_Q^2} \left(q + \frac{ q_0-\sigma_Q^2 k(t) x/\sigma_C^2}{1 + \sigma_Q^2 k^2(t)/\sigma_C^2 }\right)^2 \right. \right. \nonumber\\
		&~& \qquad \qquad \qquad \qquad \qquad \qquad \qquad \left. \left. - \frac{\left(x - q_0 k(t) \right)^2}{2\sigma_C^2 \left(1 + \sigma_Q^2 k^2(t)/\sigma_C^2\right)}\right]\right. \nonumber\\ 
		&~& \qquad +  \left. \exp\left[-\frac{ \left(1 + \sigma_Q^2 k^2(t)/\sigma_C^2 \right) }{2 \sigma_Q^2} \left(q - \frac{ q_0+\sigma_Q^2 k(t) x/\sigma_C^2}{1 + \sigma_Q^2 k^2(t)/\sigma_C^2 }\right)^2 \right. \right. \nonumber\\
		&~& \qquad \qquad \qquad \qquad \qquad \qquad \qquad \left. \left. - \frac{\left(x + q_0 k(t) \right)^2}{2\sigma_C^2 \left(1 + \sigma_Q^2 k^2(t)\right/\sigma_C^2)}\right]\right\}.\label{Pxqt}
	\end{eqnarray}
	\normalsize
	If we integrate Eq.~(\ref{Pxqt}) with respect to $q$, we get
	\begin{eqnarray}
		P_C(x,t) &=& \int dq \, P(x,q,t) \nonumber\\
		&=& \frac{1}{4\pi \sigma_C \left(1 + \sigma_Q^2 k^2(t)/\sigma_C^2\right)^{1/2}} \;  \left\{\exp\left[- \frac{\left(x - q_0 k(t) \right)^2}{2\sigma_C^2 \left(1 + \sigma_Q^2 k^2(t)/\sigma_C^2\right)}\right]\right. \nonumber\\ 
		&~& \qquad \qquad \qquad \qquad + \left. \exp\left[- \frac{\left(x + q_0 k(t) \right)^2}{2\sigma_C^2 \left(1 + \sigma_Q^2 k^2(t)/\sigma_C^2\right)}\right]\right\}.\label{AppPc}	
	\end{eqnarray}
	
	It is also possible to write the joint probability $P(x,q,t)$ in a form that is useful for determining the conditional probability $P_{Q|C}(q|x,t)$ in the limit $\sigma_C << \sigma_Q k(t)$. We first rewrite $P(x,q,t)$ in the form
	\begin{eqnarray}
		P(x,q,t) &=&  \frac{1}{4\pi}\frac{1}{\sqrt{\sigma_C^2 \left(1 + \sigma_Q^2 k^2(t)/\sigma_C^2\right)}} \frac{1}{ \sqrt{\sigma_Q^2/\left(1 + \sigma_Q^2 k^2(t)/\sigma_C^2 \right)}}\nonumber\\  
		&~&  \left\{\exp\left[-\frac{ \left(1 + \sigma_Q^2 k^2(t)/\sigma_C^2 \right) }{2 \sigma_Q^2} \left(q + \frac{ q_0-\sigma_Q^2 k(t) x/\sigma_C^2}{1 + \sigma_Q^2 k^2(t)/\sigma_C^2 }\right)^2 \right. \right. \nonumber\\  
		&~& \left. \left.  \qquad \qquad \qquad \qquad \qquad - \frac{\left(x - q_0 k(t) \right)^2}{2\sigma_C^2 \left(1 + \sigma_Q^2 k^2(t)/\sigma_C^2\right)}\right]\right. \nonumber\\ 
		&~& \qquad \qquad + \exp\left[-\frac{ \left(1 + \sigma_Q^2 k^2(t)/\sigma_C^2 \right) }{2 \sigma_Q^2} \left(q - \frac{ q_0+\sigma_Q^2 k(t) x/\sigma_C^2}{1 + \sigma_Q^2 k^2(t)/\sigma_C^2 }\right)^2 \right. \nonumber\\
		&~& \left. \left. \qquad  \qquad \qquad \qquad \qquad  - \frac{\left(x + q_0 k(t) \right)^2}{2\sigma_C^2 \left(1 + \sigma_Q^2 k^2(t)/\sigma_C^2\right)}\right]\right\}\,.\end{eqnarray}
	In the limit $\sigma_C << \sigma_Q p(t)$, we get
	\begin{eqnarray}
		P(x,q,t) &=&  \frac{1}{4\pi}\frac{1}{\sigma_C \sigma_Q } \left\{\exp\left[-\frac{ k^2(t) }{2 \sigma_C^2} \left(q - \frac{ x}{ k(t) }\right)^2  - \frac{\left(x - q_0 k(t) \right)^2}{2\sigma_Q^2 k^2(t)}\right]\right. \nonumber\\ 
		&~& \qquad \qquad + \left. \exp\left[-\frac{ k^2(t) }{2 \sigma_C^2} \left(q - \frac{ x}{ k(t) }\right)^2   - \frac{\left(x + q_0 k(t) \right)^2}{2\sigma_Q^2 k^2(t)}\right]\right\}\nonumber\\	
		&=&  \frac{k(t)}{\sqrt{2\pi}\sigma_C}\exp\left[-\frac{ k^2(t) }{2 \sigma_C^2} \left(q - \frac{ x}{ k(t) }\right)^2 \right]\nonumber\\
		&~& \qquad \qquad \frac{1}{2\sqrt{2\pi}\sigma_Q k(t)} \exp\left[- \frac{\left(x - q_0 k(t) \right)^2}{2\sigma_Q^2 k^2(t)}- \frac{\left(x + q_0 k(t) \right)^2}{2\sigma_Q^2 k^2(t)}\right] \nonumber\\ 
		&=&  \frac{k(t)}{\sqrt{2\pi}\sigma_C}\exp\left[-\frac{ k^2(t) }{2 \sigma_C^2} \left(q - \frac{ x}{ k(t) }\right)^2 \right]\nonumber\\
		&~& \qquad \qquad \frac{1}{2\sqrt{2\pi}\sigma_Q k(t)} \exp\left[- \frac{\left(x - q_0 k(t) \right)^2}{2\sigma_Q^2 k^2(t)}- \frac{\left(x + q_0 k(t) \right)^2}{2\sigma_Q^2 k^2(t)}\right]\nonumber\\ 	
		&=:&  P_{Q|C}(q|x,t)P_C(x|t).\label{AppPqgivenx}
	\end{eqnarray}
	from which $P_{Q|C}(q|x,t)$ can be read off.
	\section{Classical mixtures}
	\label{App:classicalMixtures} We consider a solution $S_{C|q}(x|q,t)$ of the Hamilton-Jacobi equation that depends on a parameter $q$ and a corresponding probability density $P_{C|q}(x|q,t)$ which, together with  $S_{C|q}(x|q,t)$, satisfies the continuity equation. One may define then classical mixtures of states with different values of $q$ in terms of the elements of the mixture $\{S_{C|q}(x|q,t),~P_{C|q}(x|q,t)\}$.
	
	\textbf{Definition:} A \textit{\textbf{classical mixture}} $\rho(x,p,t)$ is defined by introducing an additional coordinate $p$ and setting
	\begin{equation}
		\rho(x,p,t) =\int \, dq \,  P_q(q)  \, \delta\left(p-\frac{\partial  S_{C|q}(x|q,t)}{\partial x}\right) \,  P_{C|q}(x|q,t) 
	\end{equation}
	where $ P_q(q)  \ge 0$ and $\int \, dq \,  P_q(q) =1$ so that $\int dx\, dp\,\rho(x,p,t)=1$. One can show that \textit{a classical mixture is strictly equivalent to a classical phase-space distribution} \cite{HallReginatto2016}. Notice that it f
	
	\textbf{Definition:} A \textit{\textbf{classical pure state}} is a particular type of mixture which only contains a single configuration ensemble state labeled by a particular value of $q$, say $q=q_0$. It them follows that $P_q(q) =\delta(q-q_0)$ and
	\begin{equation}
		\rho^{(q=q_0)}(x,p,t) = \int \, dq \, \delta(q-q_0) \, \delta\left(p-\frac{\partial  S_{C|q}(x|q,t) }{\partial x}\right) \,  P_{C|q}(x|q,t) 
	\end{equation}
	Since there are no interference effects in classical mechanics, mixtures and pure states do not differ as strongly for classical ensembles as they do for quantum ensembles.
	
	Classical mixtures have a number of applications and they play an important role in the descriptions of measurements with a classical apparatus \cite{HallReginatto2016}, as we have already shown in Sec. \ref{sectionPointerAfterInteraction}. 
	
	The statistics of a classical ensemble may be described in  different ways using classical mixtures. We consider a free particle and show that an arbitrary mixture $\rho^{x_0}(x,p,t)$ based on particular states described by the family of solutions
	\begin{equation}\label{hpf}
		S_{C|x_0}(x|x_0,t) = m\frac{(x-x_0)^2}{2t},~~~~~P_{C|x_0}(x|x_0,t)=\frac{1}{t}F_{C|x_0}\left(\frac{x-x_0}{t}\right),
	\end{equation}
	can also be expressed as a mixture $\rho^{v}(x,p,t)$ of states described by the family of solutions
	\begin{equation}\label{sv}
		S_{C|v}(x|v,t) =-\frac{mv^2}{2}t+mvx, ~~~~~P_{C|v}(x|v,t)= F_{C|v}(x-vt).
	\end{equation}
	In both representations, $F$ are  functions that are only constrained by $P \ge 0$ and $\int dx \, P=1$. The solution of the Hamilton-Jacobi equation of Eq. (\ref{hpf}) is the Hamilton principal function, which describes a free particle that follows a trajectory that passes through $x$ and $x_0$, while that of Eq. (\ref{sv}) is the one obtained by separation of variables, which describes a free particle of velocity $v$ \cite{Synge1960}. 
	
	In the first case the mixture can be written
	\begin{eqnarray}
		\rho^{x_0}(x,p,t)
		&=& \int \, dx_0 \, P_{x_0}(x_0) \, \delta\left(p-\frac{m(x-x_0)}{t}\right)\, \frac{1}{t}F_{C|x_0}\left(\frac{x-x_0}{t}\right)\nonumber\\ 
		&=& \int \, dx_0 \, P_{x_0}(x_0) \, \frac{t}{m}\delta\left(\frac{pt}{m}-x+x_0\right)\, \frac{1}{t}F_{C|x_0}\left(\frac{x-x_0}{t}\right)\nonumber\\ 
		\nonumber\\ 
		&=& \frac{1}{m} P_{x_0}\left(x-\frac{pt}{m}\right) \,  F_{C|x_0}\left(\frac{p}{m}\right),
	\end{eqnarray}
	while in the second case it will take the form
	\begin{eqnarray}
		\rho_{v}(x,p,t)
		&=& \int \, dv \, P_{v}(v) \, \delta\left(p-mv\right)\, F_{C|v}\left(x-vt\right)\nonumber\\ 
		&=& \int \, dv \, P_{v}(v) \, \frac{1}{m}\delta\left(\frac{p}{m}-v\right)\, F_{C|v}\left(x-vt\right)\nonumber\\ 
		\nonumber\\ 
		&=& \frac{1}{m} P_{v}\left(\frac{p}{m}\right) \,  F_{C|v}\left(x-\frac{pt}{m}\right).
	\end{eqnarray}
	Since we want the two mixtures to describe equivalent statistics, $\rho^{x_0}(x,p,t)=\rho^{v}(x,p,t)$, we need to set the functions $P_{v}(*):=F_{C|x_0}(*)$ and $F_{C|v}(*):=P_{x_0}(*)$, which leads to the result
	\begin{equation}
		P_{v}\left(\frac{p}{m}\right) \,  F_{C|v}\left(x-\frac{pt}{m}\right) =P_{x_0}\left(x-\frac{pt}{m}\right) \,  F_{C|x_0}\left(\frac{p}{m}\right).
	\end{equation}
	
	Let us now specialize to the case where the first mixture includes just one state with a particular value of $x_0$, i.e. $x_0=0$, and the ensemble moves with average velocity $v'$ so that
	\begin{eqnarray}
		P_{x_0}\left(x_0\right)&=&\delta(x_0),\nonumber\\
		F_{C|x_0}\left(\frac{x}{t}\right)&=&\frac{1}{\sqrt{2\pi }\alpha} e^{-\frac{1}{2 } \left(\frac{x-v't}{\alpha t}\right)^2}=\frac{1}{\sqrt{2\pi }\alpha} e^{-\frac{1}{2 } \left[\frac{1}{\alpha}\left(\frac{x}{t}-v'\right)\right]^2}.
	\end{eqnarray}
	Then, the second mixture will be equivalent to the first mixture if
	\begin{equation}
		P_{v}\left(\frac{p}{m}\right)=\frac{1}{\sqrt{2\pi }\alpha} e^{-\frac{1}{2 } \left[\frac{1}{\alpha}\left(\frac{p}{m}-v'\right)\right]^2},~~~~~F_{C|v}\left(x-\frac{pt}{m}\right)=\delta\left(x-\frac{pt}{m}\right),
	\end{equation}
	which defines the mixture
	\begin{eqnarray}
		\rho^{v}(x,p,t)
		&=& \int \, dv \, P_{v}(v) \, \delta\left(p-\frac{\partial S_{C|v}(x|v,t)}{\partial x}\right)\, F_{C|v}\left(x-vt\right)\nonumber\\ 
		&=& \frac{1}{\sqrt{2\pi }\alpha}\int \, dv \,  e^{-\frac{1}{2 } \left(\frac{v-v'}{\alpha}\right)^2} \, \delta\left(p-\frac{\partial S_{C|v}(x|v,t)}{\partial x}\right)\, \delta\left(x-vt\right),\qquad
	\end{eqnarray}
	where we have made use of the fact that the delta function  $\delta\left(p-\frac{\partial S_{C|v}(x|v,t)}{\partial x}\right)$ forces $p=mv$. 
	Thus, the first mixture, $\rho^{x_0}(x,p,t)$, has now been expressed in terms of a new mixture $\rho^{v}(x,p,t)$ that describes particles moving with velocities $v$, where the probability of $v$ is given by a Gaussian centered at $v'$ with spread $\sigma=\alpha$. 
	
	\bibliographystyle{unsrt}
	\bibliography{version_arXiv_2}
	
\end{document}